# Laser writing of bright colours on near-percolation plasmonic reflector arrays

Alexander S. Roberts[1], Sergey M. Novikov[1], Yuanqing Yang[1], Yiting Chen[1], Sergejs Boroviks[1], Jonas Beermann[1], N. Asger Mortensen[1,2] & Sergey I. Bozhevolnyi[1,2]*

[1]*Centre for Nano Optics, University of Southern Denmark, Campusvej 55, DK-5230 Odense M, Denmark*

[2]*Danish Institute for Advanced Study, University of Southern Denmark, Campusvej 55, DK-5230 Odense M, Denmark*

**Colouration by surface nanostructuring has attracted a great deal of attention by the virtue of making use of environment-friendly recyclable materials and generating non-bleaching colours[1-8]. Recently, it was found possible to delegate the task of colour printing to laser post-processing that modifies carefully designed and fabricated nanostructures[9,10]. Here we take the next crucial step in the development of structural colour printing by dispensing with preformed nanostructures and using instead near-percolation metal films atop dielectric-metal sandwiches, i.e., near-percolation plasmonic reflector arrays. Scanning rapidly (~ 20 $\mu$m/s) across 4-nm-thin island-like gold films supported by 30-nm-thin silica layers atop 100-nm-thick gold layers with a strongly focused Ti-sapphire laser beam, while adjusting the average laser power from 1 to 10 mW, we produce bright colours varying from green to red by laser-heating-induced merging and reshaping of gold islands. Selection of strongly heated islands and their reshaping, both originating from the excitation of plasmonic resonances, are strongly influenced by the polarization direction of laser illumination, so that the colours produced are well pronounced only when viewed with the same polarization. Conversely, the laser colour writing with circular polarizations results in bright polarization-independent colour images. The fabrication procedure for near-percolation reflector arrays is exceedingly simple and scalable to mass production, while the laser-induced modification occurs inherently with the subwavelength resolution. This unique combination of remarkable features makes the approach developed for laser colour writing readily amenable for practical implementation and use in diverse applications ranging from nanoscale patterning for security marking to large-scale colour printing for decoration.**

*e-mail: seib@mci.sdu.dk

Ultrafast laser processing of materials holds important implications for both applied and fundamental research[11], including novel possibilities for post-processing and reconfiguring nanophotonic structures[12,13]. Laser processing at the nanoscale often involves resonant (local) field enhancement and photo-thermal effects, for example, to modify the morphology of individual plasmonic resonators[14] or dielectric nanoparticles (NPs)[10]. Fascinating applications range from the ultrafast delivery of heat at the nanoscale[15] to laser writing of plasmonic colours with sub-diffraction-limit resolution[9]. Laser ablation, achieved typically at high radiation fluences, can also be used for colouration of plasmonic surfaces[1], although with considerably lower spatial resolutions. Previous works on plasmonic colours[2,3] have extensively utilised progress in nanofabrication technologies to accurately pattern surfaces with carefully designed plasmonic nanostructures[4–6], using also replication techniques oriented towards mass production[7,8]. Considering a very large body of research conducted in the field of structural colours[1-10], it seems *unavoidable* that high-resolution colour printing requires high-resolution fabrication of designed nanostructures, which would result in appropriate colours either immediately[1-8] or after post-processing by laser writing[9,10].

In this work, we demonstrate a conceptually new approach to high-resolution laser printing of structural colours (Fig. 1) that eliminates the need for preformed nanostructures, using instead near-percolation metal films atop dielectric-metal sandwiches, i.e., near-percolation plasmonic reflector arrays (NPPRAs). The main idea is to take advantage of unique scattering-absorption properties of gap-surface plasmon (GSP) resonators[16,17] (that make these particularly attractive for the environment-protected colour printing[12]) and laser-induced modification of local resonance properties in near-percolation metal films[18-20]. Here, the GSP back-reflector configuration is pivotal for improving the colour vibrancy, which is generally recognized as a common challenge for plasmonic colours[10], while a very high NP surface-density in near-percolation metal films ensures high spatial resolution in laser-heating-induced NP reshaping and thereby in colour writing. Furthermore, due to inherent stability of GSP resonators with respect to surfactants, the fabricated colour prints can be protected with a transparent dielectric overlay (deposited before or after colour printing) for ambient use without destroying its colouring[5]. Finally, the NPPRA fabrication is exceedingly simple and requires only three consecutive material depositions, producing an optically thick metal back-reflector, a thin dielectric spacer and a very thin semi-continuous metal film.

The NPPRA-based laser-writing platform (Fig. 1) is amenable to a variety of modifications through the rich physics involved in thermally and locally induced NP reshaping and associated dependencies on interfacial energies, surface tensions and laser (pulse) characteristics, among others, and is thus intimately dependent on material and geometric choices as well as spatial and temporal properties of the laser. The thermally induced NP merging and reshaping can, in general, be viewed as a part of dewetting that describes the rupture of a thin (melted) film on the substrate and the formation of droplets. Thermally induced dewetting of a top metal film (as a whole) in metal-insulator-metal structures has been found suitable for fabrication of plasmonic broadband visible absorbers[21] and solar absorbers[22], replacing typically slow and expensive lithographic nano-patterning step with a straightforward and cost-effective heating procedure. Similar *global* dewetting of a silver film on glass has been used to fabricate a layer of plasmonic NPs producing dichroic colours when viewed in transmission/reflection[23] as in the famous Lycurgus cup. *Local* laser-heating-induced NP reshaping in NPPRAs introduced in our work allows one not only to realize laser colour writing with subwavelength resolution but also to achieve polarization selectivity, an interesting feature that might be useful for polarization multiplexing.

The NPPRA-based configuration investigated in detail in this work comprises 4-nm-thin (on average) gold films evaporated on 30-nm-thick silica ($SiO_2$) layers atop optically thick (100 nm) gold films supported by silicon substrates (Fig. 1). The laser colour writing is performed with a strongly focused Ti-sapphire laser (750 nm wavelength) beam with linear polarization and average powers ranging from 1 to 11 mW with the power level maintained in a closed feedback loop and controlled synchronously with beam positioning via a computer interface (see Supplementary Section 1). The top island-like gold film contains differently sized NPs that are gradually merged and reshaped due to local laser-induced heating, changing thereby the colour of reflected light (Fig. 1). Viewing through an analyser that is co-polarized with the polarization of the writing beam reveals bright and deeply saturated colours that span from yellow for the pristine sample and low-power exposures to green (for intermediate powers) and red for the largest powers (Fig. 2a). Notably, the colours are uniform even when viewed through diffraction-limited optics, indicating that the density of reshaped NPs is sufficiently high to average out local variations in the optical response on scales comparable to the diffraction limit as observed in a colour print of the University of Southern Denmark (SDU) logo (Fig. 2b) fabricated using a computerized script[5]. Comparing these images with those viewed through a cross-polarized analyser (Figs. 2c and 2d) demonstrates a high degree of polarization selectivity for colours obtained at powers up to ~ 4 mW, indicating thereby the colour range that can be accessed independently in both linear polarisations. At higher powers, the colours seen in the cross-polarised configuration start to divert from the pristine NPPRA colour, and begin to exhibit colours seen in the co-polarized configuration for lower powers. Different colours observed in the co- and cross-polarized configurations are associated with different bright-field reflectance spectra (Figs. 2e and 2f) that can in turn be mapped on a standard

CIE-1931 and linear sRGB colour spaces (Figs. 2g and 2h). The colours observed without using an analyser and the reflection spectra obtained with unpolarized light represent an intermediate case as expected (see Supplementary Section 4). In general, even though the cross talk in the images written and viewed with orthogonal polarizations is substantial, the degree of their differences is still appreciable and can further be enhanced using image processing methods and recognition techniques. We believe that this interesting and important feature might be found useful for polarization multiplexing (see Supplementary Section 4).

Local laser-heating-induced NP reshaping in NPPRAs and ensuing surface colouration introduced in our work is expected to feature subwavelength resolution because of very high densities of NPs (>> 100 $\mu m^{-2}$) typical for near-percolation metal films. Similar to the colour laser printing on preformed nanostructures[9,10], the NP reshaping requires a certain level of the writing laser power, a feature that opens a way to achieving, in general, deep subwavelength resolutions by properly adjusting the writing laser power. On the other hand, in the absence of well-defined unit cells (that one can immediately judge as being modified or not), it is difficult to quantify the achieved resolution using non-optical characterization, such as SEM imaging. Using optical images, even with high magnifications and numerical apertures, allows one to determine the resolution only up to the diffraction limit of half of the wavelength used. In this respect, the optical image shown in Fig. 2b indicates that the spatial resolution is close to or even better than the wavelength when imaging with the co-polarized (with respect to the writing polarization) illumination.

In the course of our investigations, we have used for colour printing different NPPRA configurations, i.e., having thicknesses of both spacer $SiO_2$ layers and thin island-like gold films, and found the effect of polarization-selective laser colour printing to be remarkably robust (see Supplementary Section 6). The practical applicability of colour printing schemes hinges on the robustness of the printed colours under often chemically and mechanically corrosive influences of the environment. While colours in the simple configuration with a laser-reshaped gold top layer can simply be wiped off with a light touch, the presented approach suggests various schemes of protection with a dielectric overlay due to the inherent stability of GSP resonators with respect to surfactants[5]. Thus, we have demonstrated both that the fabricated colour prints can be protected with a transparent dielectric (polymer) overlay for ambient use without destroying its colouring, and the possibility for laser colour printing on $SiO_2$-protected NPPRAs (see Supplementary Section 7).

Laser-heating induced morphological changes in near-percolation metal films have recently been considered using two-photon luminescence (TPL) imaging[18,19] and scanning (amplitude-phase resolved) near-field optical micrscopy[19] as well as bright-field optical microscopy along with electron energy-loss spectroscopy (EELS)[20]. In general, highly localized heating, which results in NP merging and reshaping upon laser light exposure, originates from the excitation of strongly enhanced and confined (dipolar) resonant surface plasmon (SP) excitations, whose existence for thin near-percolation (island-like) metal films illuminated with practically any wavelength is well documented[19,20,24-26]. The presence of an optically thick metal back-reflector changes drastically both underlying physical mechanisms involved and their optical manifestation. Thus, in the case of plasmonic metasurfaces, we established that the presence of a back reflector enhances dramatically the scattering contribution from induced magnetic dipoles[17] and opens thereby the possibility for controlling the reflected field phase within the whole $2\pi$ range[27]. In the current case of laser colour writing with NPPRAs, we have experimentally established using TPL imaging of coloured NPPRA areas intimate relationship between plasmonic resonances (see Supplementary Section 8).

Further understanding of underlying physical mechanisms involved in the proposed approach for colour printing requires constructing an adequate physical system amenable to modelling of its optical properties. To this end, we have verified the correlation between the surface morphology reconstructed using the SEM images and the optical reflectance spectra calculated using three-dimensional finite-difference time-domain (FDTD) simulations[28]. Using the SEM images of three different areas (pristine and illuminated with a linearly polarized 3.9 and 7.3 mW beam) of a 4-nm-thin gold island-like film showing three distinct colours, yellow, green and red (Fig. 3a), we calculated the reflectance spectra for the co- and cross-polarized illumination (Fig. 3b) and their colour representations (Fig. 3c). It is seen that the simulations reproduce qualitatively the experimental observations of spectra and colours, including the effect of polarization sensitivity (cf. Fig. 2e-h and Fig. 3b,c) that is apparently introduced by the melting-induced anisotropy in NP shapes[19,20]. The electric ($E_x$) and magnetic ($H_y$) field magnitude distributions (Fig. 3d) indicate the occurrence of localized SP modes (hot spots) typical for near-percolation metal films[24-26] revealed by the electric field distributions as well as relatively weakly localized GSP modes (due to their scattering by random nanostructures) revealed by the magnetic field distributions. These underlying physical mechanisms responsible for both selection of strongly heated NPs and their merging and reshaping were further revealed by simulating the absorption spectra for different material compositions based on the same random nanostructure reconstructed from the SEM image of the red-coloured surface (see Supplementary Section 2). We found the absorption spectra originate from the interband absorption in the thick bottom gold layer for short (< 500 nm) wavelengths, while the direct absorption in the top gold nanostructured film contributes primarily to the total absorption for long (> 700 nm) wavelengths. In the intermediate wavelength range (500 nm < $\lambda$ < 700 nm), the absorption due to the strongly localized SP and weakly localized GSP modes becomes dominant. Finally, a special role played by the circumstance that NP shapes are different from spherical ones for the observed polarization sensitivity[19,20] was further emphasized using simulations of reflectance spectra from regular arrays of gold ellipsoidal NPs supported by dielectric-metal sandwiches (see Supplementary Section 3).

The polarization selectivity discussed at length in the previous sections becomes detrimental when targeting conventional colour printing with prints to be viewed with ambient (unpolarized) light. In order to produce polarization-independent colour images we employed circularly polarized light for laser colour printing. We found that the corresponding colour prints look identical for all polarizations as well as for unpolarised light (see Supplementary Section 5) and even exhibit a bit brighter and more intense colours than those observed with the co-polarized writing and viewing, although the difference being admittedly small (cf. Fig 2b and Fig. 4a). Polarization insensitive printing with circularly polarized light is a nontrivial feature, when taking into account the demonstration of chiral SP multiple scattering by random surface nanostructures exploited for on-chip spectro-polarimetry[29]. In the current configuration, random NPs involved in scattering of various SP modes as discussed above are significantly smaller, and the regime of strong multiple scattering (resulting in strong localization) required for chiral sensitivity[29] is apparently not accessible. We further used the printed image for determination of the spatial resolution by superimposing optical microscopy and SEM images and highlighting thereby the correlation between structural colours and the morphology of island-like film (Fig. 4). Analysing the SEM zoom shown in Fig. 4c, one infers that the width of transition between modified (by laser writing) and pristine film surface is at the level of 400 nm, or even smaller, justifying thereby our claim of the subwavelength resolution in the considered approach for laser colour printing.

In summary, we have demonstrated the usage of near-percolation plasmonic reflector arrays for laser printing of structured colours. Scanning rapidly (~ 20 $\mu$m/s) across 4-nm-thin island-like gold films supported by 30-nm-thin silica layers atop 100-nm-thick gold layers with a strongly focused Ti-sapphire laser beam, while adjusting the average laser power from 1 to 10 mW, we have produced bright colours varying from green to red by laser-heating-induced merging and reshaping of gold islands. Both polarization sensitive and independent colour writing were demonstrated using linearly and circularly polarized light, respectively. In the case of linearly polarized writing, the difference between cross-polarized images is substantial, suggesting important perspectives for polarization multiplexing. The fabrication procedure for near-percolation reflector arrays is exceedingly simple requiring only three consecutive material depositions (to produce an optically thick metal back-reflector, a thin dielectric spacer and a very thin semi-continuous metal film) and thereby truly scalable to mass production. At the same time, the laser-induced modification occurs inherently with the subwavelength resolution, while the fabricated colour prints can be protected with a transparent dielectric overlay for ambient use without destroying its colouring. Moreover, we have also demonstrated the possibility for laser colour printing on polymer-protected near-percolation plasmonic reflector arrays, a procedure that allows one to separate the process of pristine sample fabrication and that of laser colour printing. The unique combination of aforementioned remarkable features makes the approach developed for laser colour writing readily amenable for practical implementation and use in diverse applications ranging from nanoscale patterning for security marking to large-scale colour printing for decoration.

**Methods**

Methods and any associated references are available in the online version of the paper as supplementary information.

**Acknowledgments**

S. I. B. acknowledges the European Research Council, Grant 341054 (PLAQNAP). N. A. M. is a VILLUM Investigator supported by VILLUM FONDEN (grant No. 16498). Centre for Nano Optics is financially supported by the University of Southern Denmark (SDU 2020 funding).

**Author contributions**

S. I. B. conceived the idea. A. S. R. and S. M. N. contributed equally to this work. A. S. R. and Y. C. fabricated the substrates. Y. C. did the SEM characterisation, while A. S. R. and S. M. N. performed the laser writing and optical characterisation. J. B. performed characterization by two-photon induced photoluminescence microscopy. Y. Y. conducted the analysis of underlying physical effects involved in NPPRA colouring. S. B. evaluated contributions to the reflection spectra of differently sized NPs using a semi-analytic model. S. I. B., J. B., and N. A. M. supervised the project and provided feedback on the experiments. All authors contributed to the interpretation of results and participated in the preparation of manuscript.

**Additional information**

Supplementary information is available in the online version of the paper. Reprints and permission information is available online at www.nature.com/reprints.

**Competing financial interests**

The authors declare no competing financial interests.

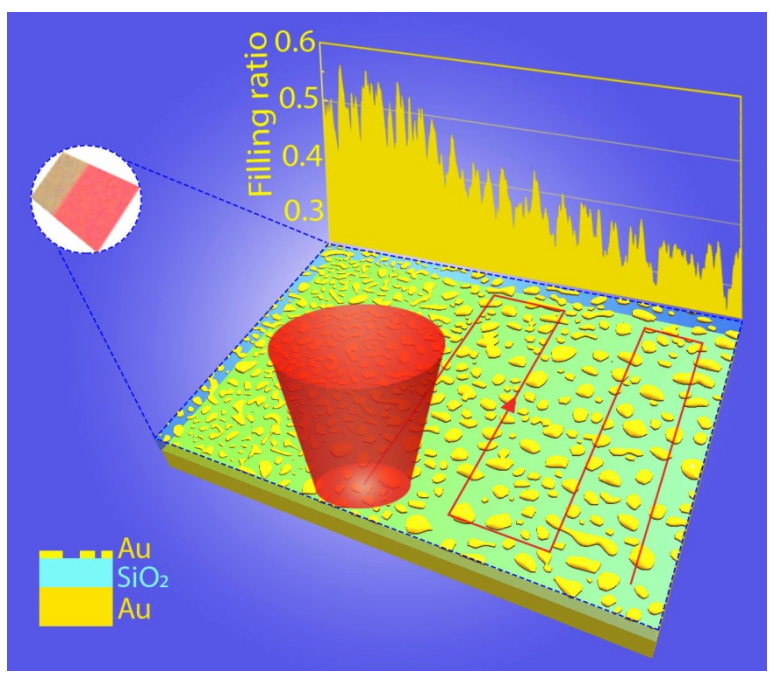

**Figure 1 | Laser colour writing on near-percolation plasmonic reflector arrays.** Schematic of laser-heating-induced reshaping of gold islands of near-percolation thin gold films atop silica-gold sandwiches. The island-like film shown represents three-dimensional rendering of an actual SEM image (1132×821 nm$^2$) taken from a sample (consisting of a 4-nm-thin gold film supported by a 30-nm-thin silica layer atop a 100-nm-thick gold layer), whose right half acquired red colour after illumination with a strongly focused 10-mW laser beam as shown on the inset to the left. Both laser beam and left inset are shown not to scale. The lateral dependence of the filling ratio was calculated from the extended (in the other direction) version of the same SEM image (1132×3621 nm$^2$), reflecting the fact that reshaping by local melting produces island in-plane shrinking and merging.

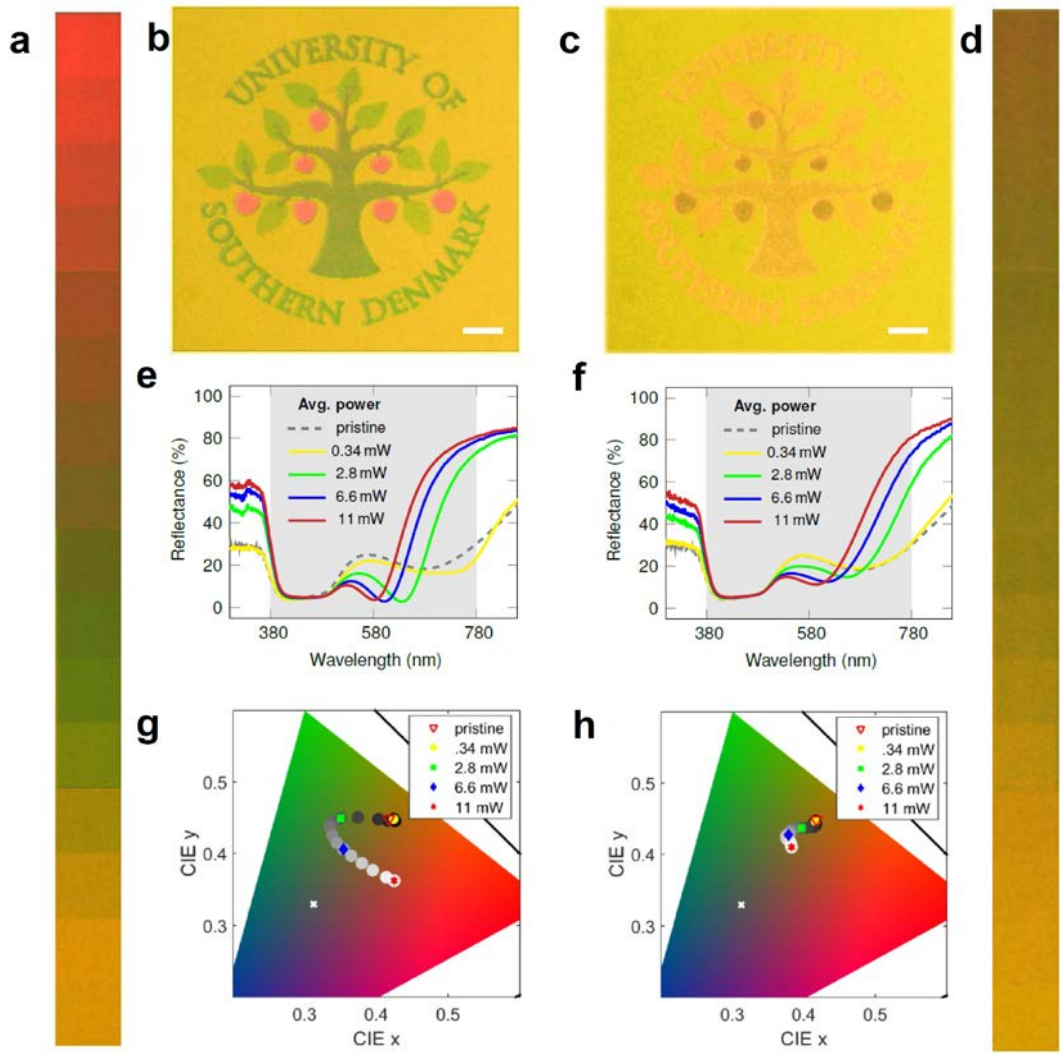

**Figure 2 | Colour and spectral analyses of anisotropic laser colour writing.** Bright-field optical microscopy images (×50 magnification, NA = 0.80) of (**a**, **d**) 15 × 15 $\mu m^2$ colour squares produced with different average laser powers increasing gradually from 0 to 11 mW and (**b**, **c**) a 90 × 90 $\mu m^2$ colour print of the SDU logo viewed with an analyser (**a**, **b**) co-polarized and (**c**, **d**) cross-polarized with respect to the polarization of writing laser beam. Scale bars: 10 $\mu$m. **e**, **f** Optical reflectance spectra and (**g**, **h**) their representations in a standard CIE-1931 and linear sRGB colour spaces corresponding to the colour squares shown in (**a**) and (**d**), respectively.

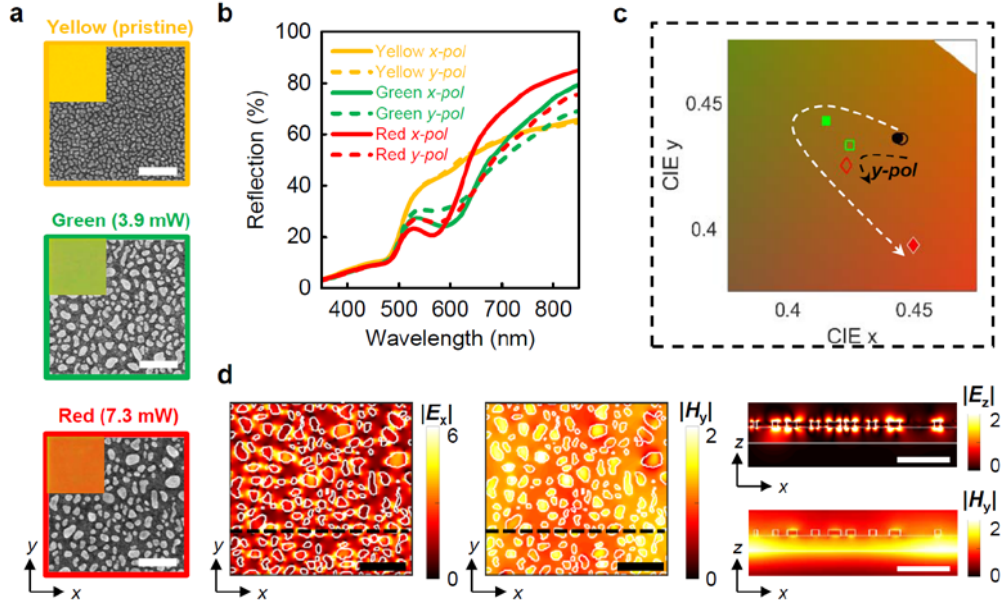

**Figure 3 | Modelling of anisotropic laser colour writing. a** SEM images of a pristine 4-nm-thin gold island-like film (yellow frame) and those exposed to the laser colour writing with an *x*-polarized 3.9 mW (green frame) and 7.3 mW (red frame) beam. Scale bars: 100 nm. The insets represent optical bright-field microscopy images of the corresponding film areas (~ 15 $\mu m^2$ in size). **b** Simulated reflectance spectra of island-like films shown in (**a**) for the co- and cross-polarized illumination (solid and dashed lines, respectively), revealing anisotropy in the optical properties of laser-modified films, and (**c**) their representations in a standard CIE-1931 and linear sRGB colour spaces. **d** Electric ($E_x$) and magnetic ($H_y$) field magnitude distributions calculated for the red-frame surface morphology (illuminated with the *x*-polarized light at 740 nm) in the middle spacer plane (left and middle panels) and in the cross section (right panels) along the dashed line indicated in the panels with plane distributions. Scale bars: 100 nm.

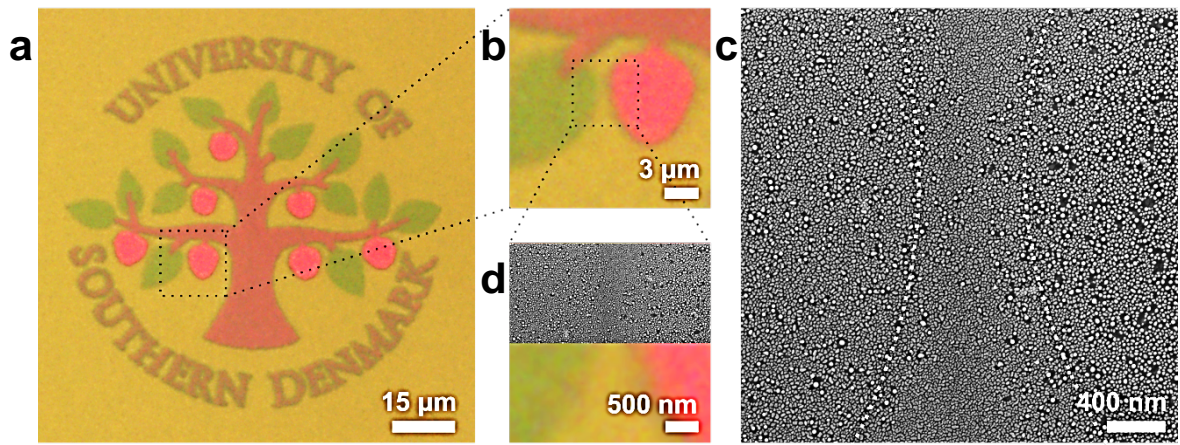

**Figure 4 | Colour writing with circularly polarized light. a** Bright-field optical microscopy (×50 magnification, NA = 0.80) image of a 90 × 90 $\mu m^2$ colour print of the SDU logo written with circularly polarized and viewed with unpolarised light. **b** Zoom (15 × 15 $\mu m^2$) of the optical microscope image shown in (**a**). **c**, **d** SEM images (3.2 × 3.2 $\mu m^2$) of the area contoured in (**b**) with the boundaries of differently coloured regions indicated with white dotted lines in (**c**) and the lower half of the SEM image being covered with the corresponding zoom of the optical microscope image shown in (**b**).

SUPPLEMENTARY INFORMATION

# Laser writing of bright colours on near-percolation plasmonic reflector arrays

Alexander S. Roberts[1], Sergey M. Novikov[1], Yuanqing Yang[1], Yiting Chen[1], Sergejs Boroviks[1], Jonas Beermann[1], N. Asger Mortensen[1,2] & Sergey I. Bozhevolnyi[1,2]∗

[1]*Centre for Nano Optics, University of Southern Denmark, Campusvej 55, DK-5230 Odense M, Denmark*

[2]*Danish Institute for Advanced Study, University of Southern Denmark, Campusvej 55, DK-5230 Odense M, Denmark*

*Correspondence to: seib@mci.sdu.dk

**This PDF file includes:**

1. Optical characterization and writing methods
2. Computer simulations of realistic near-percolation arrays
3. Coupled dipole analysis
4. Polarization selectivity, cross-talk and multiplexing
5. Polarization invariant colour printing
6. Robustness of the near-percolation plasmonic reflector array configuration
7. Printing of environmentally protected colours
8. Correlation between optical colour and two-photon luminescence images

Supplementary Figures S1- S13

Supplementary References 1- 8

## 1. Optical characterization and writing methods

Linear reflection spectroscopy is performed on a BX51 research microscope (Olympus) equipped with a halogen light source, polarizers and a fibre-coupled grating spectrometer (Ocean Optics QE65000, wavelength resolution: 1.6 nm). The reflected light is collected using an MPlanFL objective (Olympus) with ×100 magnification (NA = 0.9). Through use of a pinhole placed in a conjugated image plane, we collect spectra from an area with a diameter of ∼ 20 *μ*m. The experimental reflectance data shown in the main manuscript (Figure 2e and 2f), as well as in Supplementary Information, represent the reflection spectra calculated as the ratio $R_{str}/R_{ref}$, where $R_{str}$ is the reflection measured from colourised areas, and $R_{ref}$ is the reference obtained from a protected silver mirror (Thorlabs, PF10-03-P01) that exhibits an average reflection of 99% in the wavelength range between 350 and 1100 nm.

The two-photon photoluminescence microscopy (TPL) characterization relies on the previously described approach[1]. The experimental setup consists of a scanning optical microscope in reflection geometry built on the basis of a commercial microscope and a computer-controlled translation stage. The linearly polarized light beam from a mode-locked pulsed (pulse duration ~200 fs, repetition rate ~80 MHz) Ti-Sapphire laser (wavelength $\lambda$: 730 nm to 860 nm, $\Delta\lambda \cong 10$ nm, average power ~300 mW) is used as an illumination source at the fundamental harmonic (FH) frequency. After passing an optical isolator (to suppress back-reflection into the laser cavity), half-wave plate, polarizer, red colour filter, and wavelength selective beam splitter, the laser beam is focused on the sample surface at normal incidence with a Mitutoyo infinity-corrected long working distance objective (×100 magnification, NA = 0.70). The half-wave plate and polarizer allow accurate adjustment of the incident power. TPL radiation generated in reflection and the reflected FH beam are collected simultaneously with the same objective, separated by the wavelength selective beam splitter, directed through the appropriate filters and detected with two photomultiplier tubes (PMTs). The tube for TPL photons (within the transmission band of 350 nm to 550 nm) is connected with a photon counter giving typically only ~20 dark counts per second (cps). The FH and TPL spatial resolution at full-width-half- maximum is ~ 0.75 $\mu$m and ~ 0.35 $\mu$m, respectively, which means no individual clusters are resolved in the TPL images. In this work, we used the following scan parameters: the integration time (at one point) of 50 ms, scanning speed (between the measurement points) of 20 $\mu$m/s, and scanning step sizes of ~ 360 nm. The TPL images are obtained with the incident power of 0.3 mW. For writing, we used an incident power of 1 to 11 mW. For the reference bulk gold sample, we confirmed that the TPL signals obtained depend quadratically on the incident power. During these measurements, we kept for simplicity the excitation wavelength fixed at 740 nm, since there are no specific resonances close to this wavelength, and 740 nm is more visible and convenient to focus.

The printing of coloured images with power modulation for individual pixels is based on the setup used for TPL microscopy, with an additional electronically controllable optical attenuator (Thorlabs, EVOA800A) inserted in the beam path prior to the microscope (Figure S1). The fibre-coupled optical attenuator can be inserted and removed from the beam path by means of flip-mirrors, which guide the laser beam towards an objective (collimator) for in-coupling (out-coupling). Polarization-maintaining single-mode fibres are used to maintain control of the polarization state. For the exposure with a circularly polarized beam, an additional Glan-Taylor polarizer and subsequent achromatic quarter-wave plate (Thorlabs SAQWP05M-700) with a flat retardance in the spectral region from 300 – 1100 nm are inserted after the optical attenuator. The scanning speed of 20 $\mu$m/s used throughout for colour printing represents the largest speed, which is limited due to relatively slow operations of the variable optical attenuator (VOA) and closed-feedback loop control used for precise coordinate determination in the computer-controlled scanning stage.

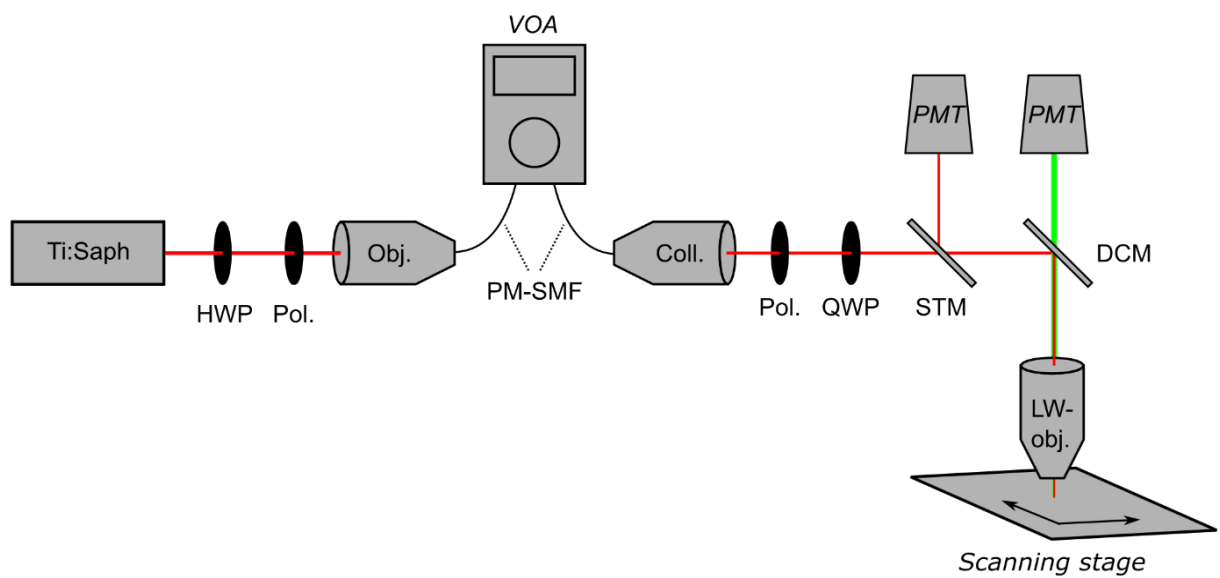

**Figure S1. Schematic of the setup used for TPL acquisition and colour printing.** The pulsed Ti:Sapphire is passed through an optical isolator (not shown), a half-wave plate (HWP) and polarizer (Pol.) in order to regulate the power, after which it is coupled to a polarization maintaining single-mode fibre (PM-SMF) via an objective (Obj.). From here, the light is directed to the variable optical attenuator (VOA) which has a computer-controlled attenuation ratio. The attenuated light passes then back to the optical path via another PM-SMF and collimated with an adjustable collimator (Coll.). A polarizer and quarter-wave plate are then used to control the polarization state

after which the light enters the microscope through a semi-transparent mirror, which is used to collect the fundamental harmonic (FH) reflection when doing two-photon luminescence measurements. A dichroic mirror (DCM) then direct the laser beam into a long working distance objective (LW-obj.) which focusses the beam onto the sample placed on a computer-controlled scanning stage.

**2. Computer simulations of realistic near-percolation arrays**

Three-dimensional finite-difference time-domain (FDTD) simulations[2] were performed with a commercially available software package (Lumerical). The permittivity of gold was taken from the experimental data determined by ellipsometry measurements on evaporated gold[3]. The structural features of the top semi-continuous gold films were directly extracted from SEM images of the films themselves. For each of the three different areas (reshaped under different laser powers) shown in Figure 3a in the main manuscript, six regions of $400 \times 400$ nm$^2$ were cropped out of the SEM image and used as multiple inputs to the FDTD simulations (Figure 3b). The height $h$ was assumed to be uniform and determined by the mass-equivalent thickness of the top gold layer $t$ (4 nm) and corresponding filling ratio $f$ as $h = t/f$. Given the different metal filling ratios of the three areas (53.1 %, 38.1 %, 32.0% for gold, green, and red areas, respectively), their heights were set as 7.5 nm, 10.5 nm, and 12.5 nm, correspondingly. Periodic boundary conditions were used in horizontal directions, and perfectly-matched layers were invoked at the top and bottom boundaries of the simulation area to absorb any transmitted or reflected waves. The overall response of each area was finally obtained by averaging the six simulated spectra (Figure S2).

To better reveal the physical origin of the spectral characteristics and associated colours of the system, we simulated four different scenarios and compare their optical responses (Figure S3). Here, the reshaped area showing red colour (Figure 3a in the main manuscript) is modelled as a representative example with the reflection and absorption spectra shown in Figure S3b. Due to the optically thick bottom gold layer, the NPPRAs can be treated as a single-port reflective system. Absorption is the only channel attenuating the reflected power and thus can well account for the features on both measured and simulated reflection spectra. Bearing this in mind, we first calculate the optical response of the percolated gold film on SiO$_2$ (Figure S3c). One observes that the direct absorption from the percolated film is almost identical to the absorption of the complete system for wavelengths > 750 nm (grey shaded area), indicating that direct absorption of the top gold film is the major contributor at long wavelengths. Moreover, the absorption peak around 570 nm of the top thin film also corresponds well to that of the complete system, indicating the importance of the percolated film structure for representing resonant features of the system.

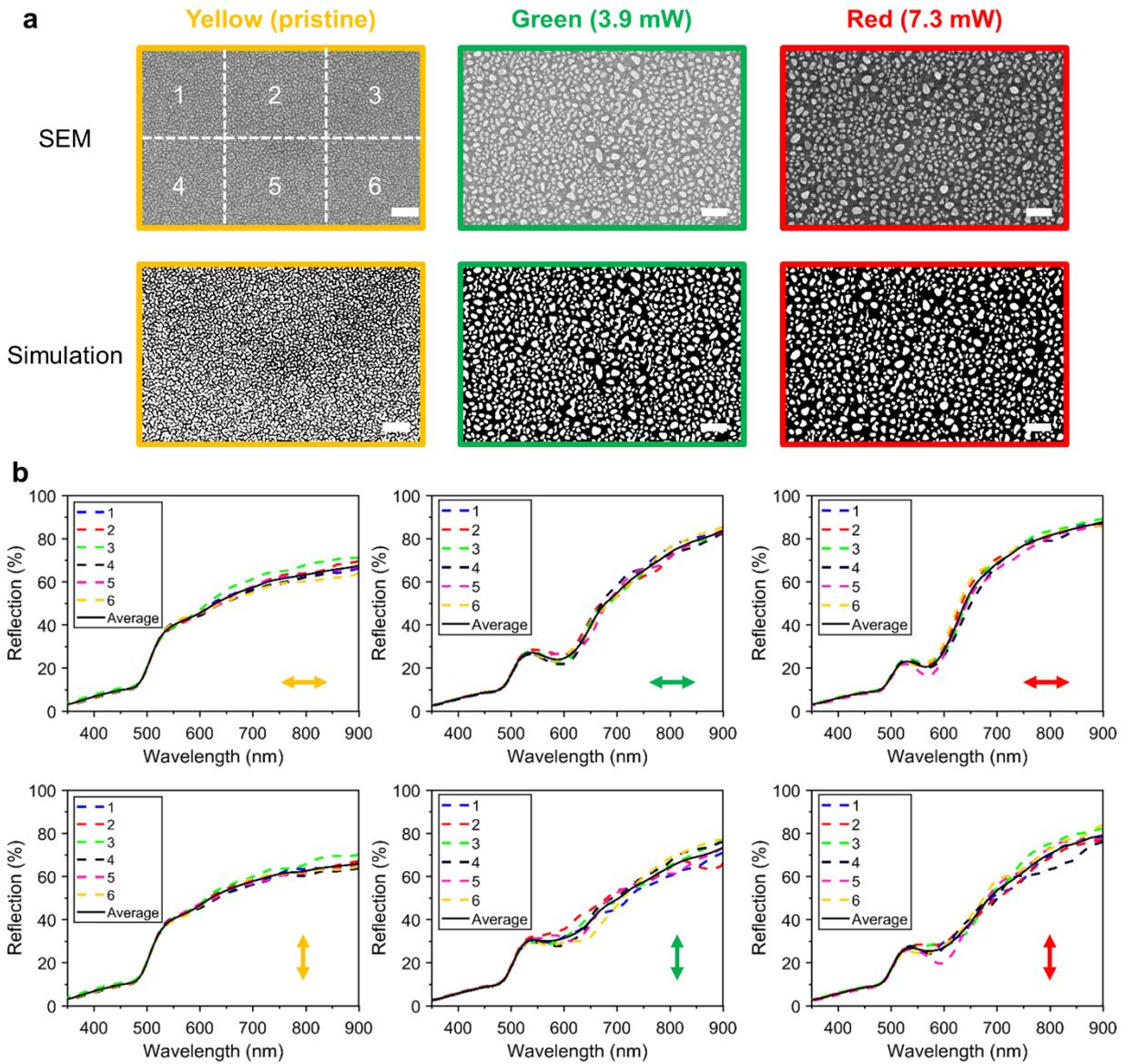

**Figure S2. Simulations of the reflection spectra.** (a) Each simulated area of the structure divided into six 400×400 nm regions, and treated numerically as described in the beginning of this section. The resulting spectra, and the average of the six spectra, are plotted in the six corresponding graphs (b), from which it emerges that the variance between the subdivided areas is small and arguably negligible. To further reduce the sensitivity towards local variations, we use the average of six 400×400 nm² area (solid black lines).

The resonant features associated with the percolated film are attributed to the occurrence of localized SP resonances of isolated gold NPs formed in the process of merging and reshaping under intense local illumination. The occurrence of SP resonances can be directly evidenced and argued for by considering electric field distributions that feature electric field hotspots (Figure 3d of the main manuscript). This finding further explains why the spectral properties (position and bandwidth) of the resonances change significantly with the laser power and exhibit a strong polarization dependence. Next, to exclude the direct absorption of the top film as discussed above, we replace the top film with a perfect electrical conductor (PEC) material while maintaining its structural features. In this case, the bottom gold layer is the only absorption source for the incoming light (Figure S3d). For short wavelengths ($\lambda < 500$ nm) the interband absorption of the bottom gold layer dominates the absorption of the complete system, whereas for longer wavelengths (500 nm < $\lambda < 700$ nm) the absorption spectrum manifests an appreciable feature of SPP dispersion, providing the evidence for the generation of SPPs propagating along the $SiO_2$-Au interface. The topmost PEC NPs, though they cannot directly give rise to any absorption, can function as efficient nanoantennae that mediate coupling to SPPs. The near-field distributions shown in Figure 3d of the main manuscript also demonstrate the existence of SP modes propagating along the bottom interface.

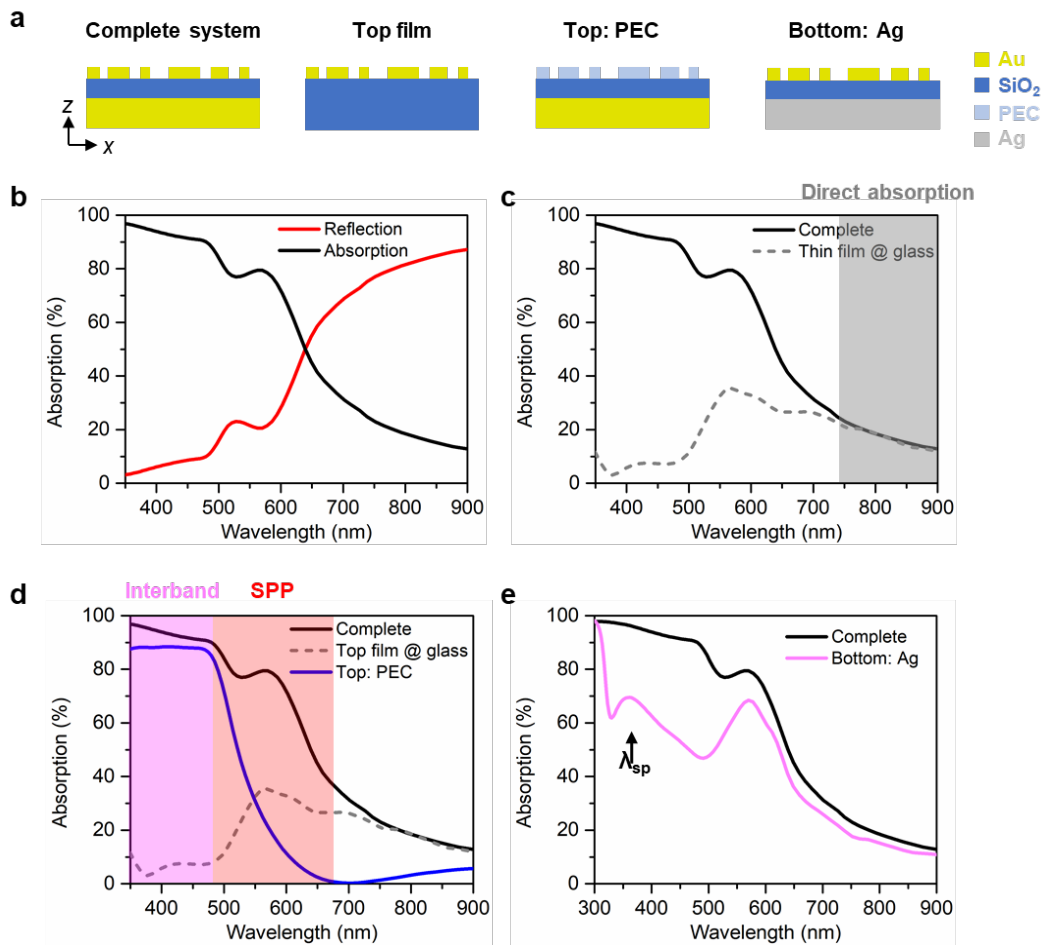

**Figure S3. Simulations of reflection spectra for different material combinations.** (a) Schematic of the considered cases used to identify different physic origins of the spectral characteristics and associated colour. The geometry of the top percolated layer is inferred from SEM images (region 1 of the area with red colour in Figure S2a) while the thicknesses of the underlying layers are $t_{SiO2}$ =

30 nm and $t_{Au}$ = 100 nm. (b) Absorption and reflection of the complete system as used for colour printing. (c) Absorption of the complete system (black, solid) and the unaltered top gold layer on a semi-infinite $SiO_2$ layer (grey, dashed). (d) Absorption for a perfect electrical conductor top layer [with similar geometry to (c)] placed on the unaltered $SiO_2$/Au substrate (blue, solid). (e) Absorption of the structure when the bottom continuous gold layer is replaced with silver (magenta, solid). The absorption of the complete system is plotted as a reference.

To further separate the two mechanisms involving the bottom gold layer, we apply a silver bottom layer instead (Figure S3e). In this case, the absorption of the silver system is dramatically smaller compared to that of the complete system at short wavelengths in the range 350 nm < $\lambda$ < 500 nm, because silver does not exhibit any interband absorption at this wavelength range, thus unambiguously verifying our explanations above. Moreover, the dispersion feature of the SP waves supported by the $SiO_2$-Ag interface and the asymptotic SP wavelength of silver (~ 350 nm) can both be easily observed and identified. In addition, we would like to note that, for the complete system, the excitation of GSP modes between the top and bottom layer (see magnetic field distributions shown in Figure 3d in the main manuscript) cannot possibly be separated and studied individually in these scenarios. Nevertheless, we can see that at certain wavelengths (e.g. the resonance peak ~570 nm in Figure S3d), the total absorption from the top film and the bottom layer is still lower than that of the complete system. Together with the magnetic field distributions shown in Figure 3d of the main manuscript, we argue that these observations provide clear evidence of the GSPs excited in our system. Overall, several distinct physical processes provide a combined contribution to the spectral features and associated colours of our system: a) for short wavelengths ($\lambda$ < 500 nm), the interband absorption from the bottom gold layer dominates the total absorption; b) for long wavelengths ($\lambda$ > 750 nm), the direct absorption from the top gold film primarily contributes to the total absorption; c) for intermediate wavelengths (500 nm < $\lambda$ < 750 nm), three mechanisms, i.e., direct absorption of the top film, SP modes propagating along $SiO_2$-Au interfaces, and GSPs between the NP layer and thick bottom layer together produce high absorption and the resonant features observed in the spectra.

**3. Coupled dipole analysis**

For a sufficiently advanced dewetting process, i.e. for colours written at high powers, the percolated film reconfigures into a film of discrete particles that can with good accuracy be described by arrays of large (relative to pristine-film features) polarisable ellipsoids using a modified long wavelength approximation (MLWA)[4]. The contribution of the shape and orientation of individual particles to the overall reflection spectra can be evaluated using transmission line theory (TLT), under the assumption that particles remain only weakly coupled[5]. In our framework involving three layers ($i$ =

1; 2; 3), one of the several physical mechanisms involved in the colour formation can be illustrated, i.e. induced dipole moments in the particles in the proximity of the metal substrate. The film of particles has subwavelength thickness ($t \ll \lambda$) and thus effectively serves to add a surface current **K** to the interface between the two homogeneous media (in this case semi-infinite vacuum with refractive index $n_1 = 1$ and a glass layer with $n_2 = 1.45$ and thickness $t_2$) in the vicinity of a semi-infinite metal reflector with $n_3 = n_{Au}$ (described by interpolated experimental data[6]). To facilitate a semi-analytic model, we consider an array of identical particles, which support an interface-averaged induced surface current[7]:

$$\mathbf{K} = -\frac{i\omega\boldsymbol{\alpha}^{\text{eff}}}{\Lambda^2}(\hat{z} \times \mathbf{E}_1)$$

where $\omega = ck_0$ is the angular frequency of the incident light, $\boldsymbol{\alpha}^{\text{eff}}$ is the effective polarizability tensor (calculated using MLWA), $\Lambda$ is the (artificial) array period, $\hat{z}$ is the normal to the surface, and $\mathbf{E}_1$ is the electric field vector. Light is incident on the array of particles normal to the surface of the homogenized film. Matching fields at the two interfaces results in the following expression for the reflection coefficient of the three-layer system:

$$r_{\nu\nu} = \frac{e^{in_2 k_0 t_2}\left(1 - i\frac{\Lambda^2(Z_1+Z_2)}{\alpha^{\text{eff}}_{\nu\nu}\omega Z_1 Z_2}\right)\frac{Z_2-Z_3}{Z_2+Z_3} - 1 - i\frac{\Lambda^2(Z_1-Z_2)}{\alpha^{\text{eff}}_{\nu\nu}\omega Z_1 Z_2}}{e^{in_2 k_0 t_2}\left(1 - i\frac{\Lambda^2(Z_1-Z_2)}{\alpha^{\text{eff}}_{\nu\nu}\omega Z_1 Z_2}\right)\frac{Z_2-Z_3}{Z_2+Z_3} - 1 - i\frac{\Lambda^2(Z_1+Z_2)}{\alpha^{\text{eff}}_{\nu\nu}\omega Z_1 Z_2}}$$

where $Z_i$ is the wave impedance and $\alpha^{\text{eff}}_{\nu\nu}$ are diagonal elements of the effective polarisability tensor with $\nu = x, y, z$ (Figure S4). Reflection spectra for 40-nm-period metasurfaces with prolate spheroids characterized by the constant volume ($5 \cdot 10^3$ nm$^3$ - average determined from experimental SEM images) but having different aspect ratios were calculated using the above formulae (Fig. S5).

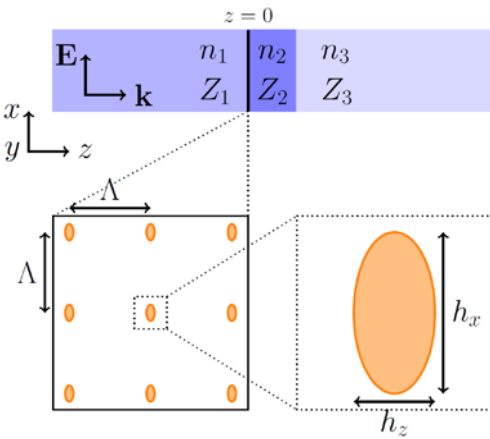

**Figure S4. Schematic of the coupled dipole model.** Bulk refractive indices $n_1$, $n_2$ and $n_3$ and associated wave impedances $Z_1$, $Z_2$ and $Z_3$ experienced by the incident light are shown (top). The quadratic array of period $\Lambda$ of coupled dipoles is positioned between layers 1 and 2, giving rise to the surface current **K** is shown bottom left. The individual ellipsoids are shown bottom right (out-of-plane dimension $h_z$ not shown).

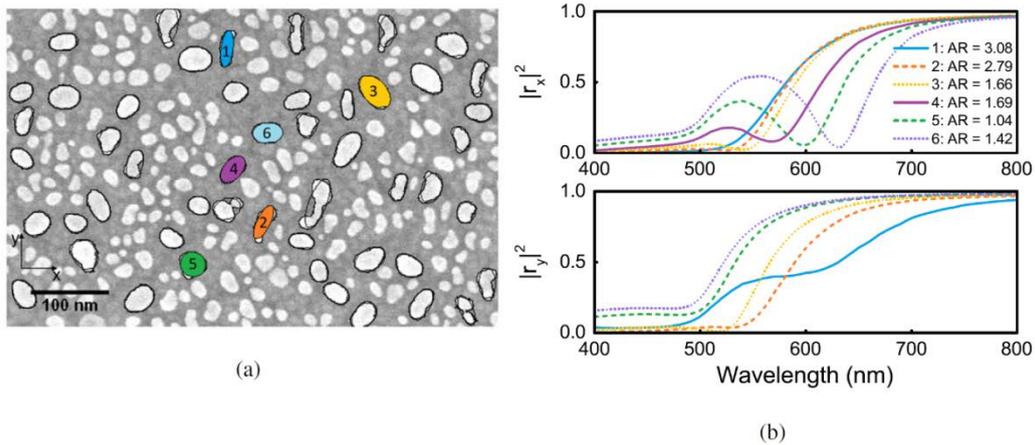

**Figure S5. Reflection spectra from ellipsoidal NPs of different sizes.** (a) SEM image used for the evaluation of the resemblance of coupled-dipole arrays of ellipsoids fitted to particle shapes as viewed in the SEM. Six particles (1-6, colour coded) with areas larger than 500 nm$^2$ are chosen. (b) NPs chosen in (a) are simulated as an array of coupled dipoles in the framework of transmission-line theory under the MLWA for an array period of 40 nm under *x*- (upper) and *y*- polarized (lower) illumination. The aspect ratio (AR) between major and minor axes of the ellipsoidal particles is indicated in the legend (see Figure S4 for the model schematic).

## 4. Polarization selectivity, cross-talk and multiplexing

Strongly polarization-selective colour-writing can be achieved using linearly polarized light, and yields colours with strong polarization anisotropy for all colours, with bright saturated colours resulting from viewing under co-polarized illumination - including bright and dark greens, grey as well as bright and dark reds - and only weakly saturated areas appearing in the cross-polarization. It is possible to achieve even brightly saturated red in co-polarized illumination while maintaining very low colour saturation for the cross-polarized illumination. This fact is particularly noteworthy since the thermal diffusion by necessity introduces the onset of polarization cross-talk at some power threshold, while Figure S6 demonstrates that substrates can be engineered to reveal the full colour palette before the onset of strong polarization crosstalk due to thermal diffusion. We next investigated the extent to which colours can be written independently in the two linear polarization states. We chose the highest powers of each horizontal line in Figure S6 (1.3, 4.5, 7.5 & 11 mW) and twice expose a square area consisting of four lines with the chosen powers in each polarization, where both the exposure pattern and the polarization is rotated between the two exposures (Figure S7). Thus, the powers increase from top to bottom (*x*-pol) and left to right (*y*-pol). This yields a diagonally symmetric pattern when viewed under unpolarized light. When inserting a polarizer aligned along either direction, the pattern exposed in that particular polarization reveals itself - with only minimal visible colour variation resulting from the orthogonally polarized exposure.

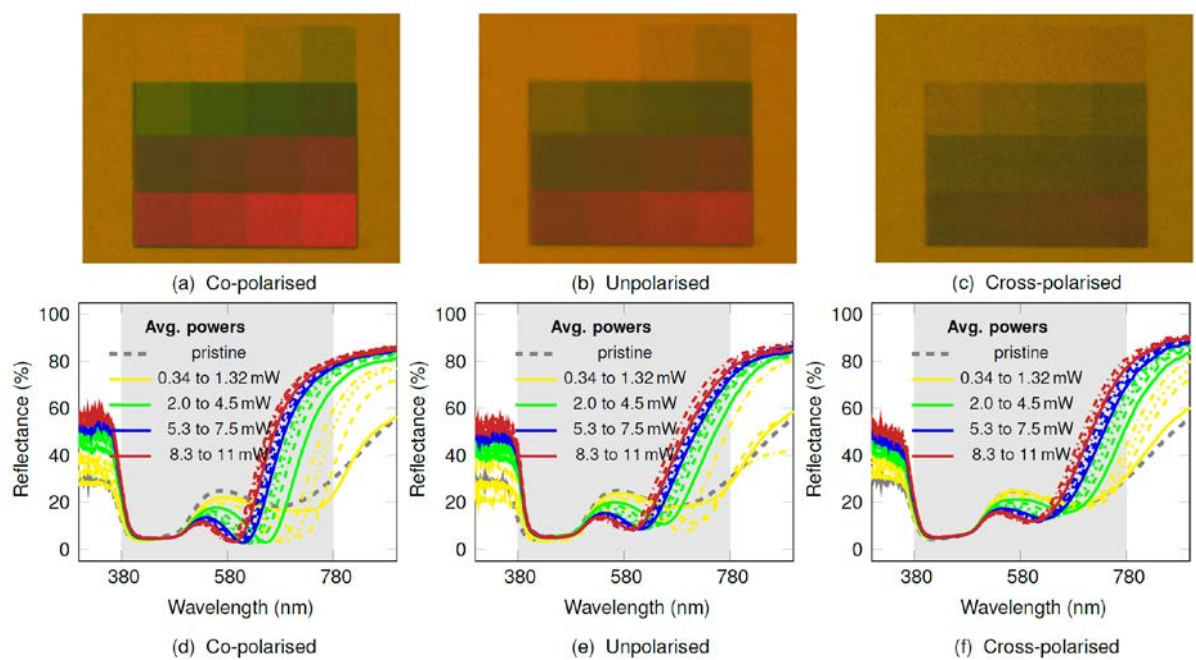

**Figure S6. Polarization selectivity.** (a) - (c) Microscopic bright-field images of 16 areas (15×15 $\mu$m$^2$ each) exposed on an Au (100 nm) - SiO$_2$ (25 nm) - Au (4 nm) substrate at increasing powers from the upper left to the lower right corner, illuminated with light (a) co-polarized, (b) unpolarized and (c) cross-polarized with respect to the writing polarization. (d) - (f) Spectral reflectance of the areas in (a) - (c), under correspondingly polarized illumination.

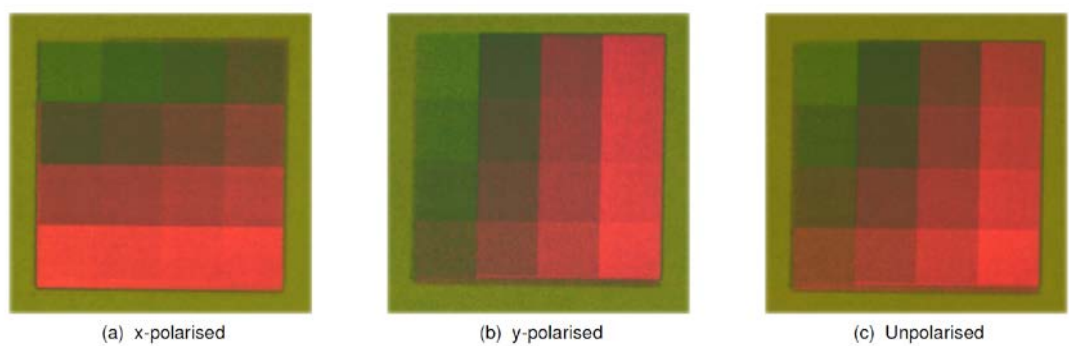

**Figure S7. Polarization cross-talk.** (a) - (c) Microscopic images of 16 areas (15×15 $\mu$m$^2$ each) exposed with *x*-polarized (horizontal) at increasing powers from the top to the bottom, and with *y*-polarized (vertical) light at increasing powers from left to right. Viewed in (a) *x*-polarized, (b) *y*-polarized and (c) unpolarized. The *x*-polarized exposure was written first.

An illustration of two different images exposed in each polarization is seen in Figure S8, where we have used local pedestrian traffic light images of Hans Christian Andersen. Figure S8 (b, c) show the green and red light images exposed in *y*- and *x*-polarization, respectively, while Figure S8 (a, d) show the intended silhouettes. It emerges clearly that each HCA silhouette stands out clearly when viewed under the intended polarization, even though the cross-polarized silhouette admittedly is still faintly visible due to the polarization cross-talk. As might be expected from Figure S7, the red (high-power) exposure leaks more strongly into the orthogonal polarization than does the green (low power) exposure.

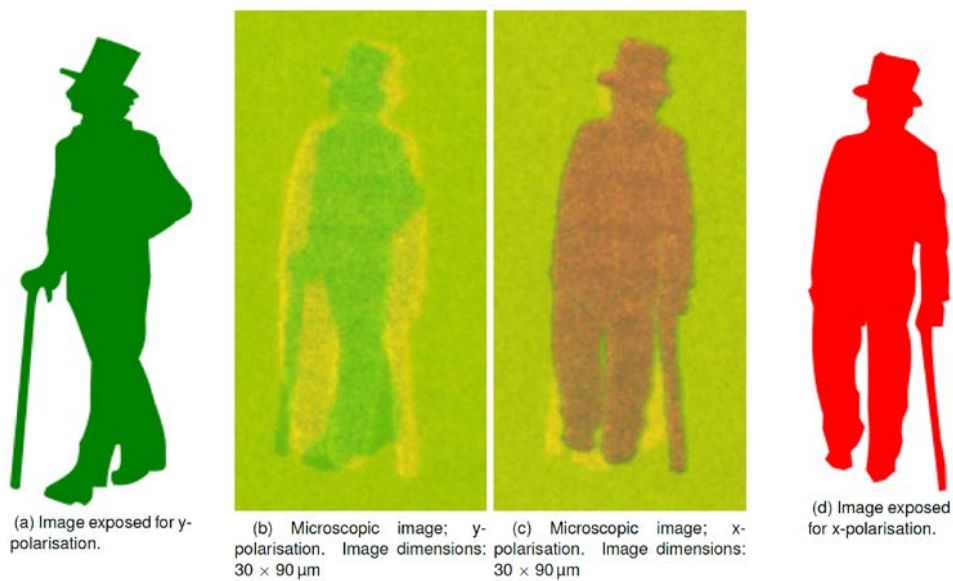

**Figure S8. Polarization multiplexing.** (a) *y*-polarized exposure pattern of a walking H.C. Andersen figure and (b) resulting microscopic bright-field image under *y*-polarized illumination. (c) Standing HCA traffic light figure viewed and exposed with *x*-polarized light. (d) Exposure pattern used for the *x*-polarized writing at the same place as that used for the *y*-polarized writing.

**5. Polarization invariant colour printing**

With the aim of achieving images that are invariant under any, linear or circular, polarization state – and for unpolarized and even partially polarized viewing - we modify our experimental setup with an additional quarter-wave plate and polarizer, enabling the exposure with a circularly polarized beam. This makes the resulting image fully invariant under both linear and circular polarizations (Figure S9). It is clearly seen that no noteworthy changes (slight homogenous variations in the background and SDU logo colours are due to variations in polarization properties of the dichroic mirror used as a beam splitter, see Figure S1) occur when changing between unpolarized illumination and the four investigated polarization states, clearly demonstrating that polarization invariant prints can indeed be obtained in a *single* scan when using a circularly polarized beam for printing colour images.

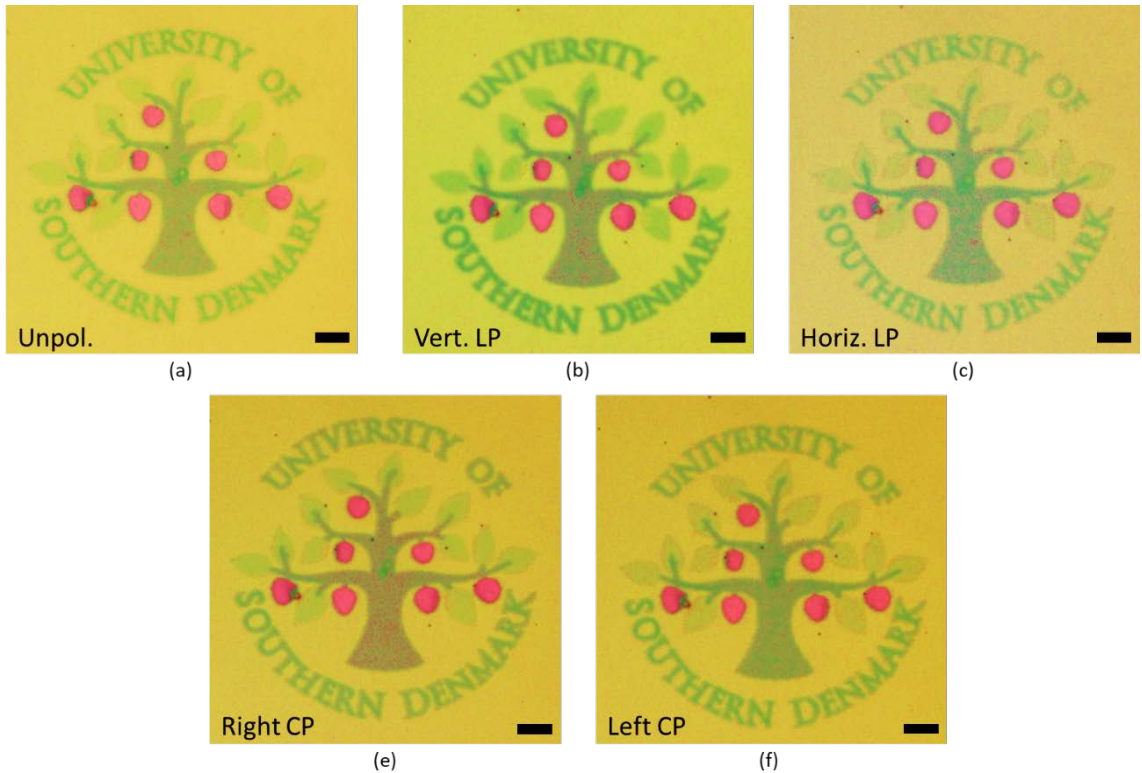

**Figure S9. Polarization invariant colour printing.** Image exposed with a circularly polarized beam. Image viewed under unpolarized illumination (a), vertical (b) and horizontal (c) linear polarization, and right (e) and left (f) circularly polarized light, demonstrating fully polarization independent behaviour. The imperfections in the image result from SEM images taken prior to the investigation of polarization invariance. Scale bars: 10 $\mu$m.

## 6. Robustness of the near-percolation plasmonic reflector array configuration

In the course of our studies, we have tried different thicknesses of both spacer $SiO_2$ layers and thin island-like gold films that were also deposited under different conditions and using different evaporation chambers and deposition techniques. Without going into detail of all these exploratory experiments, we would like to emphasize that the effect of polarization-selective laser colour printing was seen to be remarkably robust, as it has been found with all investigated configurations, albeit with somewhat different colour palettes and colour vividness (as illustrated in Figure S10 below). It should be mentioned that the parameter space of the proposed approach is enormous as it includes also various surface treatments that might strongly influence the morphology of near-percolated metal films.

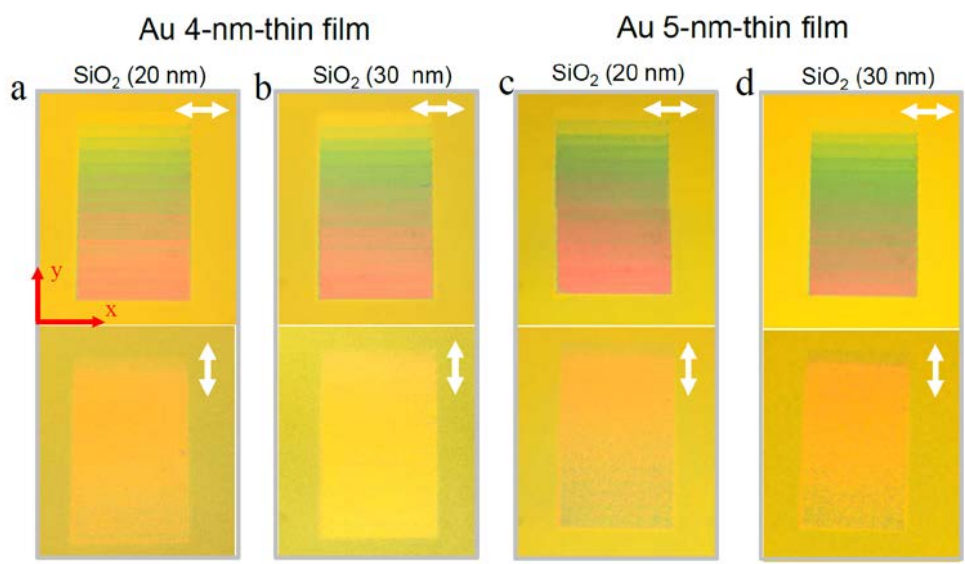

**Figure S10. Robustness of polarization-selective laser colour printing.** Optical bright-field microscopy images of (50×5 $\mu m^2$) rectangular stripes exposed with increasing (from top to bottom) powers (from 0 to 10 mW) of the *x*-polarized writing laser beam and viewed with *x*- and *y*-polarized illumination (top and bottom rows, respectively) for 4 different (in the $SiO_2$ spacer and gold thicknesses as indicated) samples.

## 7. Printing of environmentally protected colours

The practical applicability of colour printing schemes hinges on the robustness of the printed colours under the often chemically and mechanically corrosive influence of the environment. While colours in the simple configuration with a percolated top layer can simply be wiped off with a light touch, the technique nevertheless lends itself to various schemes of protection via a dielectric overlay, which, intriguingly, can be applied in either pre- or post-processing of the substrate. With the aim of showing that both colours and polarization sensitivity of the exposure are preserved when applying a PMMA protection layer in post-processing, we cover the sample from Figure S7 with roughly 230 nm of PMMA, which includes baking step for 180 seconds at 180°C. While this protection introduces obvious changes to the colours (Figure S11), as the PMMA layer acts as an antireflection stack, the resulting colours are perceived not too dissimilar to the initial ones, with a range including greens and reds. The polarization dependence is not influenced.

While the application of dielectric layers in post-processing is feasible for some applications, such as colouration of consumer products, the technique is entirely impractical for in-situ printing applications, where the printable substrate should be protected both prior, under and after printing. This creates the need to encapsulate the (close to) percolated gold layer at the fabrication stage. We demonstrate that a sample can indeed be covered with a protective layer of silicon dioxide prior to the laser writing – again with no loss of saturation or polarization selectivity (Figure S12). Both techniques result in prints that can be handled and touched with bare hands with no visible effect (not shown). It deserves mentioning that, while 20 nm of $SiO_2$ is not a robust protection layer that can withstand robustly mechanical handling, a simple increase of the layer thickness can increase the robustness significantly.

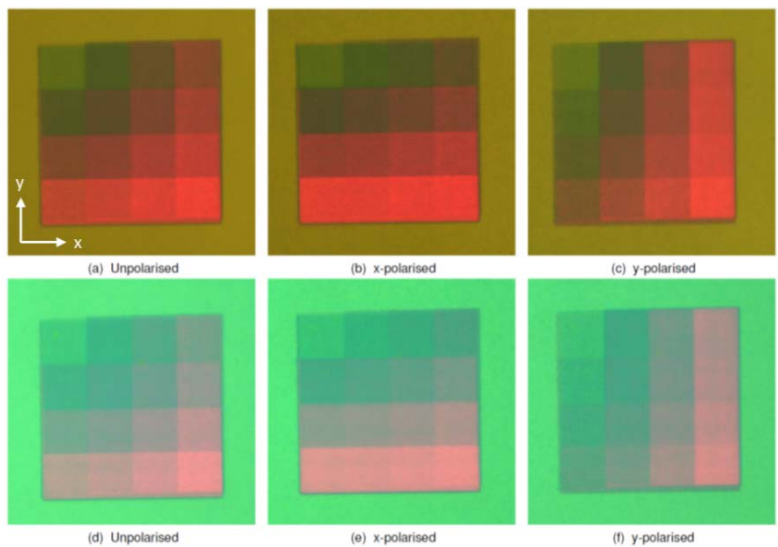

**Figure S11. Protection of printed colours.** Protection of printed colours by addition of a dielectric PMMA layer (approximately 230 nm) layer post-printing (d-f). (a-c) Polarization multiplexed scans identical to Figure S7 before PMMA deposition. (d-f) Polarization multiplexed scans identical to Figure S7 after PMMA deposition, showing that both colour saturation and polarization selectivity survive the spin-coating of a 230-nm-thick PMMA layer as well as subsequent baking at 180°C.

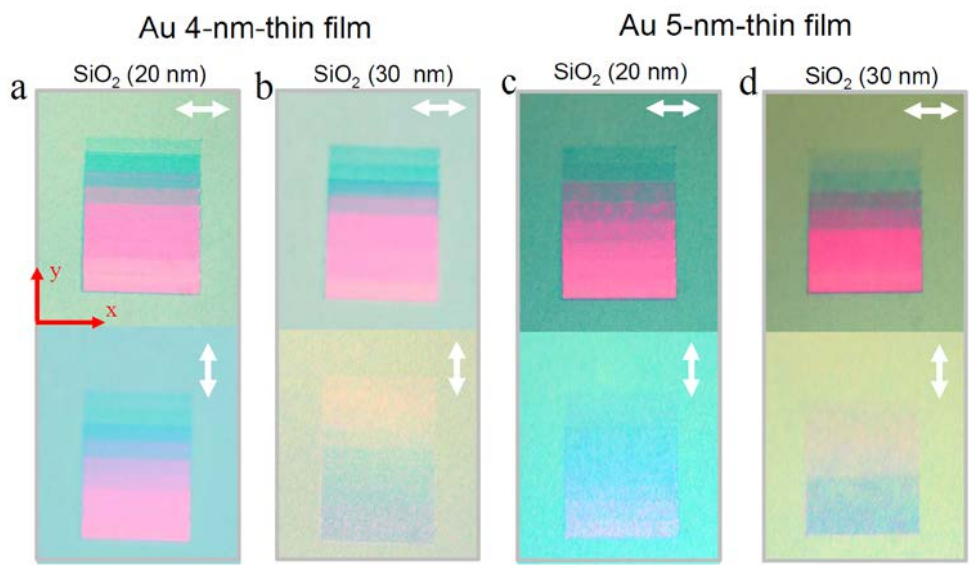

**Figure S12. Colour printing on protected near-percolated gold films.** Optical bright-field microscopy images of (50×5 $\mu m^2$) rectangular stripes exposed with increasing (from top to bottom)

powers (from 0 to 10 mW) of the *x*-polarized writing laser beam and viewed with *x*- and *y*-polarized illumination (top and bottom rows, respectively) for 4 different (in the SiO$_2$ spacer and gold thicknesses as indicated) samples that were covered with a 20-nm-thick SiO$_2$ protection film before the laser writing. Different colouration of protected pristine films compared to the previously investigated (unprotected) films (cf. Figure S10) is related to the influence of the protection layer.

## 8. Correlation between optical colour and two-photon luminescence images

We have also investigated the correlation between colour printing as observed with optical bright-field microscopy and two-photon luminescence (TPL) imaging by exposing two (5×5 $\mu$m$^2$) square areas with different powers of incident light being *x*- and, slightly above, *y*-polarized, resulting in 4 coloured areas located close to each other (Figure S13).

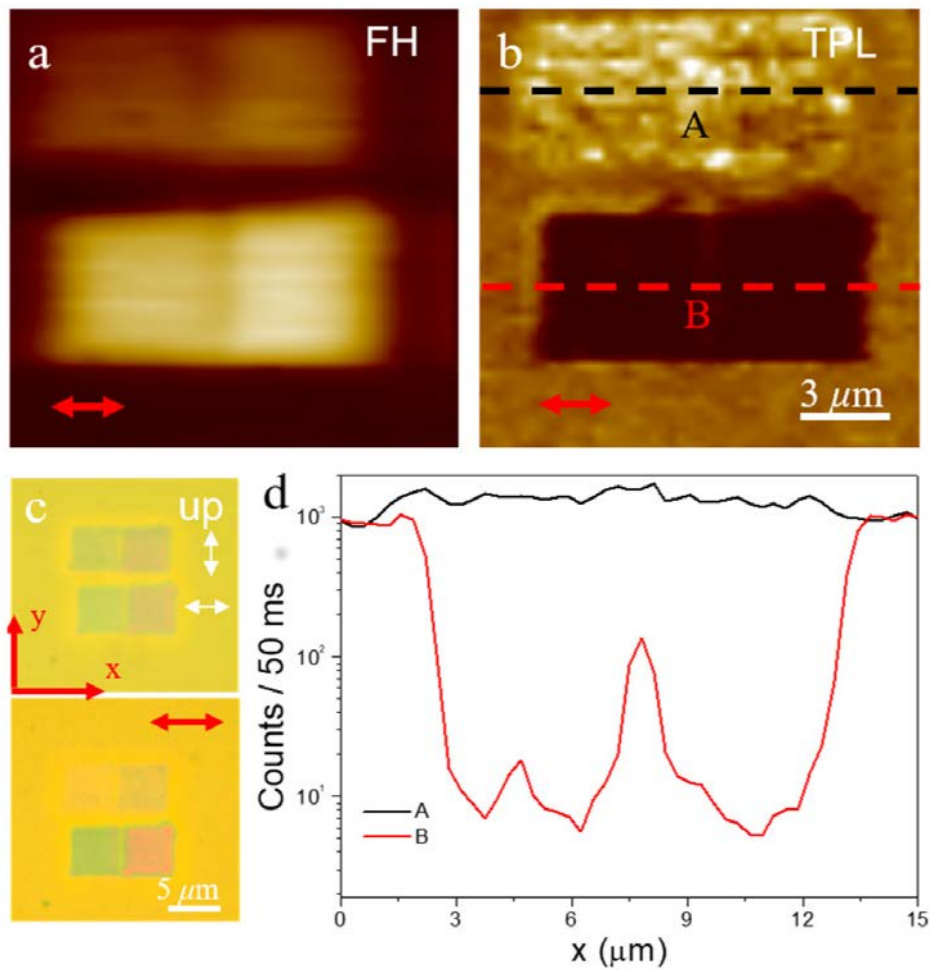

**Figure S13. Correlation between optical colour and TPL images.** (a) Fundamental harmonic and (b) TPL images of a 4-nm-thin gold film on a 30-nm-thin SiO$_2$ spacer supported by a 100-nm-thick gold film after being subjected to laser colour printing with different powers and polarizations, with *x*- and *y*-polarized light as shown in (c) by white arrows. (c) Optical bright-field microscopy

images obtained under the unpolarized (top) and *x*-polarized (bottom) illumination. (d) Averaged TPL signal obtained with the *x*-polarized illumination along the cross sections taken at the positions indicated by black and red dashed lines numbered in (b).

It is seen that, while the optical bright-field images look essentially identical for the unpolarized illumination, the polarization sensitivity in both optical and TPL images obtained with the *x*-polarized illumination is equally pronounced. This is understandable, since the optical image colouration is associated with the excitation of plasmonic resonances, and the TPL images are also correlated with local field enhancements at both the illumination and detection wavelength[8]. Rather large FH (a) and very low TPL (b) signals obtained with the *x*-polarized illumination for two lower areas indicate that the *x*-polarized writing destroys (*x*-polarized) resonances at 740 nm, while enhancing (*x*-polarized) scattering for shorter wavelengths, leaving *y*-polarized resonances intact.

**Supplementary References**